\documentclass[aps,prl,showpacs,showkeys,noeprint,superscriptaddress,secnumarabic,amssymb,nobibnotes]{revtex4-2}
\usepackage{amsmath}
\usepackage{amssymb}
\usepackage{graphicx}
\usepackage{url,hyperref}
\usepackage[usenames,dvipsnames]{xcolor}
\hypersetup{colorlinks=true, linkcolor=BrickRed, urlcolor=blue!50!black, citecolor=blue!50!black}
\usepackage[capitalize]{cleveref}
\usepackage{cleveref}
\crefrangelabelformat{equation}{(#3#1#4)$-$(#5#2#6)}

\usepackage{verbatim}

\begin{document}
%
\title{Anomalous temperature dependence in phase transitions via ballistic coalescence}
\author{Nalina Vadakkayil}
\affiliation{Theoretical Sciences Unit and School of Advanced Materials, Jawaharlal Nehru Centre for Advanced Scientific Research, Jakkur P.O., Bangalore 560064, India}
\affiliation{Engineering Mechanics Unit, Jawaharlal Nehru centre for Advanced Scientific Research, Bangalore -
560064, India}
\author{Sutapa Roy}
\affiliation{Theoretical Sciences Unit and School of Advanced Materials, Jawaharlal Nehru Centre for Advanced Scientific Research, Jakkur P.O., Bangalore 560064, India}
\affiliation{Department of Physics, Birla Institute of Technology and Science, Pilani, Hyderabad Campus, Jawahar Nagar, Kapra Mandal, Medchal District, Telangana 500078, India}
\author{Jiarul Midya}
\affiliation{Theoretical Sciences Unit and School of Advanced Materials, Jawaharlal Nehru Centre for Advanced Scientific Research, Jakkur P.O., Bangalore 560064, India}
\affiliation{School of Basic Sciences, Indian Institute of Technology  Bhubaneswar, Argul, Khordha, Odisha 752050, India}
\author{Subir K. Das}
\email{das@jncasr.ac.in}
\affiliation{Theoretical Sciences Unit and School of Advanced Materials, Jawaharlal Nehru Centre for Advanced Scientific Research, Jakkur P.O., Bangalore 560064, India}
\date{\today}
\begin{abstract}
We study kinetics of phase transitions within a single component Lennard-Jones model. For low enough particle densities disconnected clusters form that can move ballistically in an inviscid vapor background. The clusters undergo sticky collisions, thereby forming larger aggregates, which can be of fractal nature at ultra cold temperatures. As the temperature is varied, the exponent of the algebraic growth of average cluster mass changes, exhibiting pronounced nonmonotonic character. We capture this anomalous behavior, in two and three space dimensions, within a ballistic aggregation theory. We show that the scope of the theory is much broader than the typically considered case where cluster motions are uncorrelated. An analysis of the theory shows that ballistic aggregation can even occur exponentially fast. This is in sharp contrast with the conventional algebraic picture, regarding passive matter phase transitions. We discuss scenario where such explosive growth can be realized. In addition, our results are widely relevant in understanding structure and growth in aerosols, cosmic dust and other aggregation processes.
\end{abstract}
\maketitle
\section{Introduction}\label{intro}
Following quenches inside the coexistence or ordered region of a phase diagram, homogeneous systems become unstable to fluctuations and evolve towards new equilibrium~\cite{ral_jones,onuki,bray,binder_book}. Such phase transition kinetics, that initiates via formation of domains of like particles, have been studied extensively in contexts like para-to-ferromagnetic transitions, phase separation in solid binary mixtures, and vapor–liquid transitions~\cite{onuki, bray,binder_book,puri,binder1977theory,siggia,langer1971theory,san1985phase,ahmad2010kinetics,furukawa1987turbulent,lookman,datt2015,majumder2011universality,davis2025kinetics}. During the evolution processes, domain growths in these systems obey algebraic rise \cite{bray,puri,binder_book,binder1977theory,tanaka1995,tanaka1996coarsening,tanaka1997droplet,majumder2011universality,siggia,binder_stauffer,furukawa1985effect,midya2017droplet,lookman,datt2015,bera_skd,davis2025kinetics,langer1971theory,san1985phase,ahmad2010kinetics,furukawa1987turbulent,roy2012nucleation,midya2020kinetics}. Despite each of the above mentioned cases belonging to separate nonequilibrium universality class, the primary characteristics of kinetics remain robust against variation in key thermodynamic parameters. For example, in a given case the growth can be described by a unique exponent,  irrespective of the quench temperature~\cite{bray}. 
Here we present a case for which such temperature independence is replaced by striking anomalies. 

Disconnected cluster morphology can appear during phase separation if the overall particle density is low \cite{lifshitz,wagner,binder_stauffer,furukawa1985effect,roy2012nucleation,shimizu,huo2003hydrodynamic,midya2017droplet,bera_skd,paul2021clusters,midya2017prl,roy2013dynamics}. Additional complexities, for structure as well as dynamics, may occur if solid-like clusters of one phase move in the background of another \cite{paul2021clusters,midya2017prl}. For such situations, via hydrodynamics preserving molecular dynamics simulations \cite{frenkel}, we identify, in realistic Lennard-Jones systems \cite{frenkel}, sharp variations in growth exponent with the change of temperature. These anomalies are marked by unusual nonmonotonicity, presenting pronounced peaks.

For low enough temperatures, in space dimensions two and three, we observe ballistic motion of clusters that can as well be of strong fractal nature~\cite{midya2017prl}. Therefore, to understand the anomalous kinetics, we use a ballistic aggregation theory, by incorporating in it the information on fractality~\cite{carnevale,trizac2003correlations,ulrich2009dilute,midya2017prl,pathak2014inhomogeneous}. The theory reproduces the simulation results quite accurately. We show that the predictions of the theory remain accurate much beyond the traditionally considered situation where cluster motions are uncorrelated. Furthermore, the theory suggests that ballistic aggregation can even be exponentially rapid. This is in sharp contrast with the standard expectation of algebraic scaling in passive matter transitions. Our results are relevant for understanding structure and growth in a large class of systems, including cosmic dust, aerosols, and clouds~\cite{chandrasekhar1943stochastic,pnas_brilliantov}. We discuss physical situation where the exponential growth can be realized.
\section{Model and Methods}\label{model_methods}
Our model systems consist of particles of diameter $\sigma$, with densities $\rho = 0.04$ and $0.03$, in space dimensions $d=3$ and $d=2$, respectively. The particles interact with each other via a truncated, shifted and force-corrected Lennard-Jones (LJ) potential \cite{majumder2011universality,roy2012nucleation}:
\begin{equation}\label{eq:potential}
 U(r) = u(r)-u(r_c)-(r-r_c) \left(\frac{du}{dr}\right)_{r=r_c},
\end{equation}
where
\begin{equation}\label{eq:lj}
 u(r)=4\varepsilon \left[\left(\frac{\sigma}{r}\right)^{12} - \left(\frac{\sigma}{r}\right)^{6} \right],
\end{equation}
with $\varepsilon$ and $r_c$ ($=2.5\sigma$) being, respectively, the interaction strength and the cut-off distance. The critical values of density ($\rho_c$) and temperature ($T_c$) for the vapor-liquid transition within this model are estimated to be \cite{midya2017droplet,midya2017finite} $0.37$ and $0.41 \varepsilon/k_B$ in $d=2$ and $0.316$ and $ 0.939 \varepsilon/k_B$ in $d=3$, $k_B$ being the Boltzmann constant. Based on the space dimensions, we choose square or cubic boxes of linear dimensions $L = 2048$ and $128$, respectively, unless stated otherwise, with periodic boundary conditions. We carried out molecular dynamics (MD) simulations in the canonical (NVT) ensemble by implementing the Verlet velocity algorithm \cite{allen,frenkel}. The temperature was controlled via a Nos\'e-Hoover thermostat \cite{nose,hoover}. In our study, homogeneous systems were quenched to different final temperatures, $T_f$, that fall inside the coexistence region. 

For the chosen densities, the morphologies are made of discrete clusters. 
To calculate the average mass, we first identified these clusters. The particles are considered to belong to the same cluster or droplet \cite{roy2013dynamics}, if their distances from other particles fall within a radius $R  = 1.2 \sigma$. If the number of particles within a cluster exceeds a threshold number, viz., $50$, we include this for the calculation of the average mass. At a given time, $M(t)$ is obtained as
\begin{equation}
    M(t) = \frac{1}{N_d}\sum_{i=1}^{N_d}M_{d_{i}},
\end{equation}
where $N_d$ is the number of clusters and $M_{d_{i}}$ is the mass of the $i^{th}$ cluster. 

All results presented in the following are averaged over runs with a minimum of $10$ initial configurations in $d=2$ and $20$ initial configurations in $d=3$. The time in our simulations was measured in units of $\tau = (m\sigma^2/\varepsilon)^{1/2}$. The integration time step, $\Delta t$, was varied between $0.005\tau$ and $0.01\tau$, for obtaining the solution of the dynamical equations. We set $m$, $\sigma$, $\varepsilon$, and $k_B$ at unity. Parts of the results were obtained using LAMMPS~\cite{lammps}.
\section{Results}\label{results}
%
\begin{figure}
\centering
  \includegraphics*[width=0.47\textwidth]{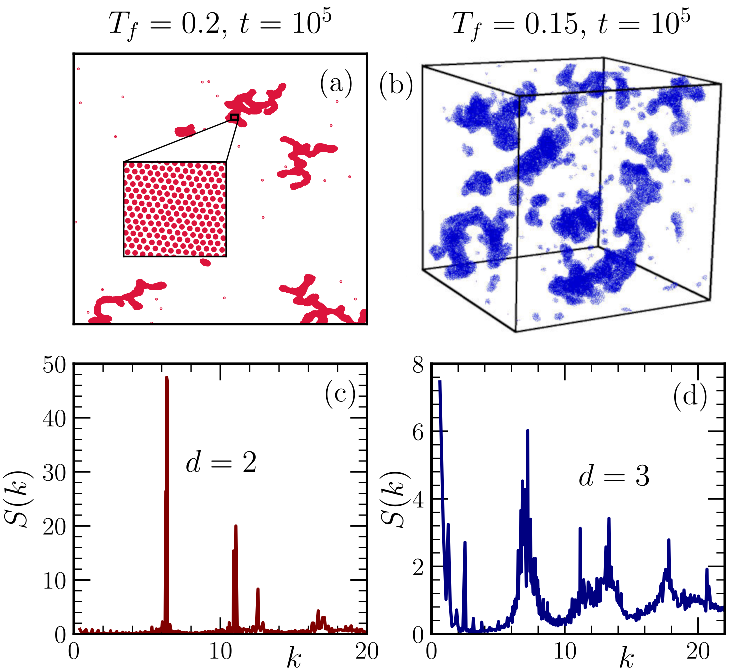}
  \caption{(a) A late time snapshot taken during the evolution following a quench of a $2D$ homogeneous configuration to $T_f = 0.2$. The linear dimension of the considered system size is $L = 1024$, of which only an enlarged region of $800\times800$ is shown for clarity. A part of a cluster is zoomed to show internal structure. (b) Late time snapshot following a quench to $T_f = 0.15$ in $d = 3$. (c) The structure factor calculated for the zoomed portion in part (a). (d) The structure factor for a part of a cluster in $d=3$.}
\label{fig:snapshots}
\end{figure}
We begin by providing the structural information in Fig. \ref{fig:snapshots}. In parts (a) and (b), we show snapshots from the evolutions of homogeneous configurations in $d=2$ and $3$, respectively. Disconnected morphology, with ramified clusters, following quenches to $T_f = 0.2$ and $0.15$, in respective dimensions, can be appreciated. In the 2D case, a part of a cluster is blown up to show regular solid-like arrangement of particles. This is true for the 3D case as well. This feature is quantitatively demonstrated in Figs. \ref{fig:snapshots}(c) and (d). There we plot $S(k)$, the structure factor, as a function of wave vector $k$. It is defined as  \cite{hansen}
\begin{equation}
    S(k) =  \frac{1}{N_{p}}\Big \langle \sum\limits_{i,j=1}^{N_p} e^{i\vec{k}.\vec{r} }\Big\rangle\,,
\end{equation}
where $N_p$ is the number of particles in a given cluster, $\vec{r}  = \vec{r}_i - \vec{r}_{j}$, and $\vec{r}_{i(j)}$ is the position vector of $i(j)$th particle. The sharp peaks confirm the solid state nature. At much higher temperatures, say, at $T_f = 0.5$, in $d = 3$, the structure within the clusters have no periodicity in particle arrangements. In such liquid-like situations one observes spherical clusters.

\begin{figure}
\centering
  \includegraphics*[width=0.47\textwidth]{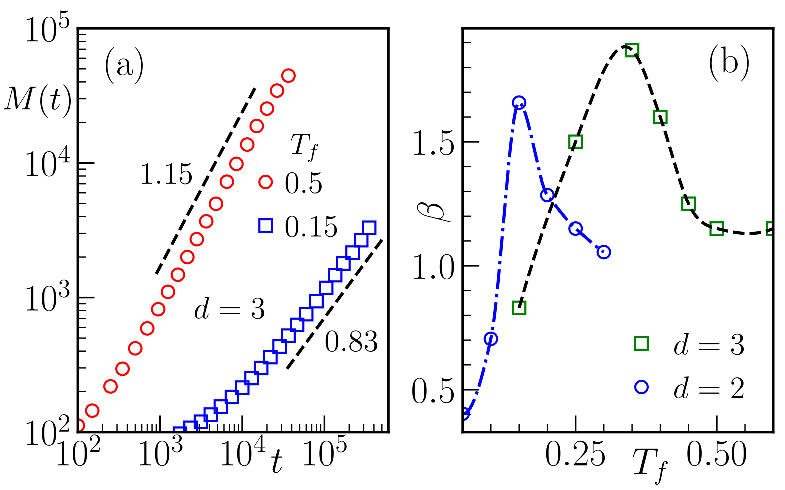}
  \caption{
  (a) Log-log plots of the average mass of the clusters, $M(t)$, as a function of time, for two different temperatures $T_f = 0.5$ and $0.15$, in $d = 3$. The dashed lines denote power laws with exponents ($\beta$) mentioned next to the lines. (b) Plots of $\beta$ versus $T_f$, for $d = 2$ and $3$. The dashed lines are interpolations, serving as guides to the eye.}
\label{fig:mass}
\end{figure}

We now focus on the growth behavior of these clusters.
In Fig. \ref{fig:mass} (a), we show the time dependence of the average mass, $M(t)$, in $d=3$. We have included data from two final temperatures. Steady power-law rise over long times can be appreciated. This allows us to estimate values of $\beta$, in 
\begin{equation}
    M \sim t^\beta\,,
\end{equation}
quite accurately. The difference between the asymptotic values of $\beta$ appears quite high, between the two cases. This implies a strong dependence of the coarsening process on the quench temperatures. To better appreciate this, we extract $\beta$ for a large range of $T_f$ values. In part (b), we show this as a function of $T_f$, for $d = 3$ as well as $ d = 2$. Irrespective of the space dimensionality, sharp nonmonotonic variation is visible, with pronounced maxima. In passive matter systems, typically it is understood that growth rate is independent of temperature, except for lowering at very low temperature, owing to dynamical arrest with metastability. In contrast, we here have a fascinating anomaly in the growth dynamics! 

Disconnected domain morphology is a property of phase separation with low overall density or asymmetric composition in binary mixtures~\cite{lifshitz,wagner,binder_stauffer,furukawa1985effect,roy2012nucleation,shimizu,huo2003hydrodynamic,midya2017droplet,das2023hydrodynamic,bera_skd,paul2021clusters,midya2017prl,roy2013dynamics}. In addition to the temperature independence mentioned above, in solid binary mixtures, irrespective of symmetric or asymmetric compositions, the growth occurs via Lifshitz-Slyozov-Wagner particle diffusion (LSW) mechanism \cite{lifshitz,wagner}. There a power-law exponent, $1/3$, for growth of average cluster size ($\ell$) is not only independent of the temperature, it also remains same regardless of space dimensions. However, in the present case, it is possible that the droplets can move, in the vapor background, and undergo collisions. In that case, the LSW mechanism will be inadequate to explain the dynamics. In fact, in fluid environment, coalescence mechanisms may become dominant due to significant movements of droplets or clusters \cite{siggia,binder_stauffer,furukawa1985effect,das2023hydrodynamic,aarts2005hydrodynamics,pawar2012arrested}. This general mechanism, for diffusive motion of the droplets, is referred to as the Binder-Stauffer mechanism, which may apply to growth with morphology having liquid as well as solid clusters. The rate of change in droplet density $n$ in this mechanism is written as \cite{binder_stauffer} $
    dn/dt = - B n^2 \,,$
where $B$ ($>0$) is a constant. This provides $n \sim 1/t$, which, combined with $n\sim 1/\ell^d$, leads to a dimension dependent power-law growth: $\ell(t) \sim t^{1/d}$. However, unlike our simulation observations, the exponent here is temperature independent.

\begin{figure}
\centering
  \includegraphics*[width=0.48\textwidth]{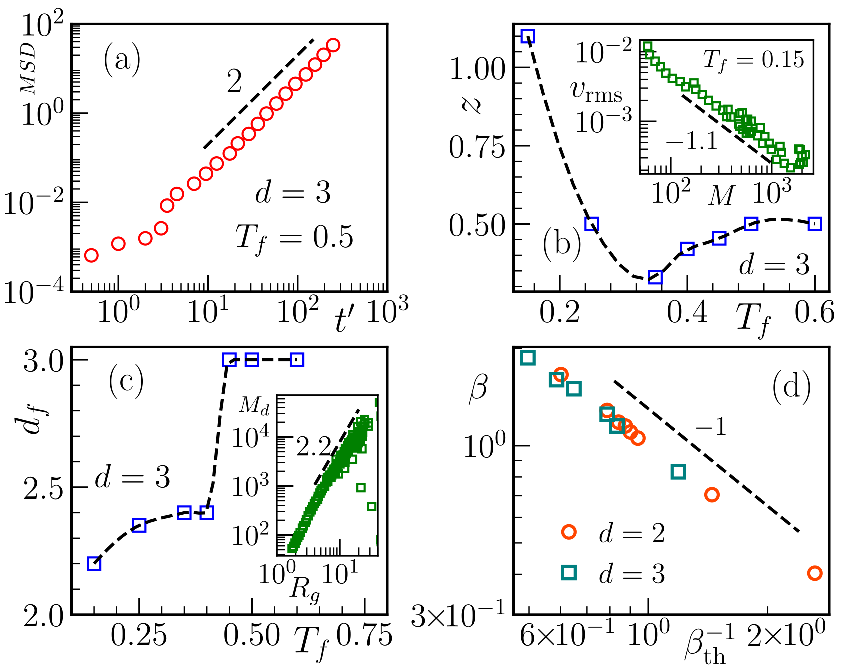}
  \caption{(a) Log-log plot of the mean-squared displacement of clusters, as a function of translated time, $t'$, for $T_f = 0.5$. The dashed line denotes a power-law with the exponent $2$. (b) Plot of $z$ as a function of $T_f$. The dashed, interpolated line, is a guide to the eye. Inset: Log-log plot of $v_{\rm{rms}}$ versus $M$ for $T_f = 0.15$. The dashed line denotes a power-law with the exponent $-1.1$. (c) Plot of $d_f$ as a function of $T_f$, with a dashed guiding line. Inset: Log-log plot of the cluster mass, $M_d$, versus $R_g$, for $t = 10^5$ and $T_f = 0.15$. The dashed line denotes a power-law with the exponent mentioned next to it. These results are for $d = 3$. (d) Plot of $\beta$ vs $\beta_{\rm{th}}^{-1}$ ($=[d_f(z+1)+1-d]/d_f$), $\beta_{\rm{th}}$ being the corresponding theoretical value, for $d =2$ and $3$. The dashed line represents a power-law with an exponent $-1$.}
\label{fig:msd_rmsv}
\end{figure}
If the background fluid phase is of low density, the clusters can move ballistically. In Fig. \ref{fig:msd_rmsv}(a) we present a representative plot for mean-squared-displacement \cite{allen,hansen}, calculated as $\langle |\vec{r}_{\rm{cm}}(t) - \vec{r}_{\rm{cm}}(0)|^2\rangle$, for a quench to $T_f = 0.5$, in $d=3$. In the above expression $\vec{r}_{\rm{cm}}$ is the position vector of the center of mass of a given cluster. Clearly, ballistic motion, even at a reasonably high $T_f$, is evident, from the consistency of the data with a behavior $\sim t^2$. Note that at this $T_f$, the underlying cluster phase is liquid. Inspired by the observation in Fig. \ref{fig:msd_rmsv} (a), we proceed to verify the relevance of ballistic-aggregation (BA) theory \cite{carnevale,trizac1995dynamic}, over a large range of $T_f$, covering both solid and liquid clusters. 

If the underlying cluster phase is fluid, after a coalescence event the product droplet quickly gains spherical shape, a requirement of energy minimization. However, if the underlying phase of the clusters is solid, there can be complexity. This scenario is true for sufficiently low temperature at which the time to relax to a symmetrical shape, following the formation of a product cluster, can be quite long. An uneven competition between time scales, in the latter scenario, between collision time and post collisional shape relaxation process, can produce fractality in the shape of clusters~\cite{midya2017prl}. This can be appreciated from Fig. \ref{fig:snapshots}.

The rate of change of the cluster density, $n$, within the BA theory is written as \cite{midya2017prl,carnevale,trizac2003correlations,ulrich2009dilute,paul2017ballistic,trizac1995dynamic},
\begin{equation}\label{eq:bta}
    \frac{dn}{dt} = -\textrm{`Collision cross section'} \times \langle v_{rel} \rangle \times n^2 \,.
\end{equation}
For uncorrelated motion of the clusters, $\langle v_{rel} \rangle$, the average relative velocity, is the same as $v_{\rm{rms}}$, the root-mean-squared velocity. The latter has a mass dependence as $\sim M^{-z}$, with \cite{carnevale} $z = 1/2$. Application of the theory has been primarily restricted to the close vicinity of this value of $z$. However, depending on $T_f$, significant deviation can occur from $1/2$, as seen in Fig. \ref{fig:msd_rmsv}(b). In the inset, we have shown how this exponent is estimated for the quench temperature $T_f = 0.15$. To quantify the collision cross section it is important to calculate $d_f$, the fractal dimensionality.  We show a plot of $d_f$, as a function of $T_f$, in Fig. \ref{fig:msd_rmsv}(c). The procedure \cite{midya2017prl} of estimating $d_f$, exploiting the relation $M \sim R_g^{d_f}$, $R_g$ being the radius of gyration, is demonstrated in the inset. A sharp change of $d_f$ between $T_f = 0.4$ and $0.45$ is due to change in the underlying droplet phase, implying, below $T_f = 0.45$ the clusters are at solid phase. This fact we have verified via the calculation of $S(k)$, as well as via the estimation of principal peak heights of radial distribution function. 

For compact droplets or clusters, the collision cross-section can be written as $M^{\frac{d-1}{d}}$. However, for fractal structures, this dependence is modified \cite{midya2017prl,ulrich2009dilute} to $M^{\frac{d-1}{d_f}}$. Here note that $n \sim M^{-1}$. Invoking these in Eq. \eqref{eq:bta}, one obtains \cite{ulrich2009dilute,midya2017prl}
\begin{equation}\label{eq:mass}
    M(t) \sim t^{\beta_{\rm{th}}}, \,\, {\rm{with}}\,\, \beta_{\rm{th}} =  \frac{d_f}{d_f(z + 1) +1-d}\,,
\end{equation}
where $\beta_{\rm{th}}$ is the BA theoretical counterpart of $\beta$. 
In part (d) of Fig. \ref{fig:msd_rmsv} we show the simulation data for $\beta$, against the inverse of $\beta_{\rm{th}}$, obtained by incorporating the values of $z$, $d_f$, and $d$. The data sets from both dimensions nicely overlap with each other. The appearance on the log-log scale, with a power-law having exponent $-1$, confirms the accuracy of the theoretical expression. This also extends the applicability of the theory significantly beyond $z = 1/2$, i.e., even when the cluster motions are reasonably correlated. 

\begin{figure}
\centering
\includegraphics*[width=0.47\textwidth]{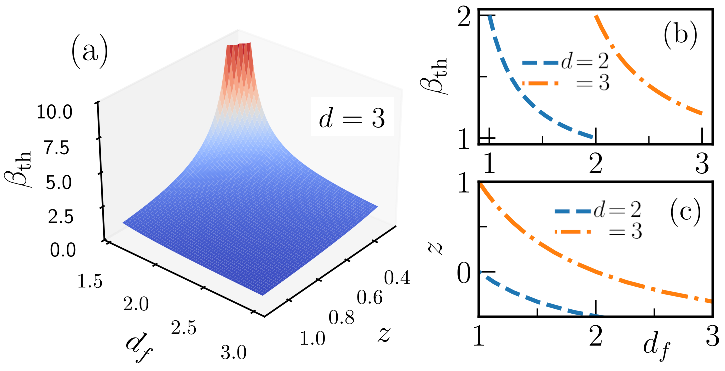}
\caption{(a) A surface plot showing variation of $\beta_{\rm{th}}$ with $d_f$ and $z$, for $d = 3$. The color code implies increasing values of $\beta_{\rm{th}}$ from blue to red. (b) Plots of $\beta_{\rm{th}}$ versus $d_f$, for $d = 2$ and $3$, when $z = 1/2$. (c) Plots of $z$ versus $d_f$, for $d = 2$ and $3$, corresponding to the exponential growth.}
\label{fig:theoretical}
\end{figure}

The appearance of the peaks in $\beta$ in Fig. \ref{fig:mass}(b) provides an impression of possible divergence of the growth exponent. Though a singularity does not appear in our simulations, the possibility can be appreciated from the examination of the expression in Eq. \eqref{eq:mass}: $\beta \to \infty$ as $d_f (z+1) \to d-1$, with the understanding that $d_f$ is finite. To elucidate this, in Fig. \ref{fig:theoretical}(a) we present a surface plot by showing the variation of $\beta_{\rm{th}}$, when $d_f$ and $z$ are changed, for $d=3$. Such large values of $\beta_{\rm{th}}$, however, are not seen in our simulations. As temperature increases beyond the peak location, see Fig. \ref{fig:mass} (b), the conditions change in such a way that the clusters become more compact, i.e., $d_f \to d$ and velocities get randomized so that $z \to 1/2$. This keeps the system away from the regime of extraordinarily large $\beta$ values. 

\begin{figure}
\centering
  \includegraphics*[width=0.45\textwidth]{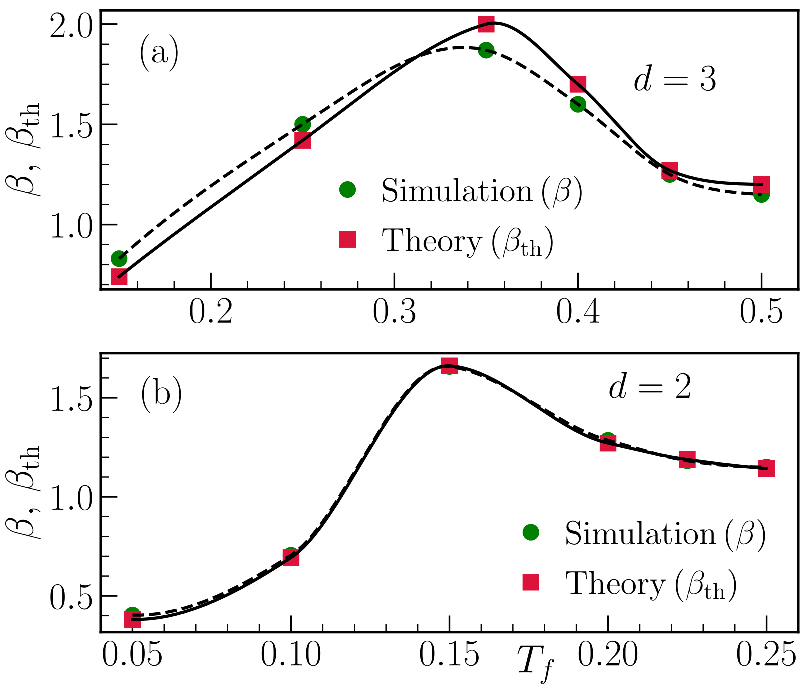}
  \caption{(a) Plot of $\beta$ as a function of $T_f$, obtained directly from the power-law behavior of the simulation data of $M(t)$, along with the values of $\beta_{\rm{th}}$, calculated by using Eq. \eqref{eq:mass}, in $d=3$. (b) Same as (a) but for $d=2$. The solid and dotted lines are guides to the eye.}
\label{fig:th_sim}
\end{figure}

In Fig. \ref{fig:theoretical}(b) we plot $\beta_{\rm{th}}$, as a function of $d_f$, for only the specific case $z= 1/2$. For this value of $z$, the divergences occur at $d_f = 4/3$ when $d=3$ and at $d_f =2/3$ for $d=2$. Here it should be noted that the divergence here essentially means an exponential growth of mass. This is a significant departure from the literature expectation of algebraic growth during phase  transitions. This scenario can be further realized separately by solving Eq. \eqref{eq:bta} for combinations of $d_f $ and $z$ that provide $dM/M \sim dt$, say, for the above quoted values of $4/3$ and $2/3$, with $z = 1/2$, in $d = 2$ and $3$. This provides $M \sim \exp(At)$, where $A$ is a positive constant. The requirement of $d_f  = 2/3$, in the 2D case, implies clusters with voids that will dissociate during movement. A value of $4/3$, in $d= 3$, on the other hand, though implies significantly filamental structure, is physically realizable, say, in dipolar systems \cite{rovigatti2012structural,sakai2025dipolar,singh2025self}. 
In these systems, particles can arrange themselves to provide even linear cluster shape, having a very low fractal dimension and thus, extremely fast growth, other environmental condition supporting. For perfectly linear structures, $d_f = 1$. In that case one requires $z=1$ for an exponential growth. As seen in our study, this is a physically realizable possibility. In fact, for many other physically realizable combinations of $d_f$ and $z$ explosive growths should be possible.

In Fig. \ref{fig:theoretical}(c) we provide loci of exponential growth possibility on $z$ vs $d_f$ plane for both $d = 2$ and $3$. Note here that the physically realizable values of $z$ are limited \cite{trizac2003correlations} to $z \geq 0$. For the present case of LJ systems, it is an open question whether an exponential growth is possible. The interactions among the particles are different here from the dipolar case. Furthermore, temperature variation brings in varying competition between the two relaxation times, as well as changes the vapor phase properties, keeping the window of ballistic aggregation somewhat restricted. Nevertheless, an exercise with the variation in both $T_f$ and $\rho$ will be useful, though computationally demanding.

Finally, direct comparison of the simulation data for $\beta$, over a wide variation of $T_f$, with the corresponding estimates of $\beta_{\rm{th}}$ are provided in Fig. \ref{fig:th_sim}. While for $d = 2$ the comparison is nearly perfect, the theory captures the behavior extremely well in $d = 3$ as well. This again confirms the breadth of applicability of the theory. Whether an exponential rate is possible or not, these results are already remarkable. Furthermore, the theory's ability to capture the anomalies at quantitative level is very exciting.

\section{Conclusion}
We have studied phase transitions in a single component Lennard-Jones system in space dimensions $d = 2$ and $3$. The growth exponents for the disconnected morphology, originating due to low overall density, show anomalous variation with the change of temperature, exhibiting pronounced peaks. Such anomalies arise due to interesting interplay between time scales associated with the cluster movements and relaxation following post collision compound cluster formation~\cite{midya2017prl}. To interpret phase separation with low overall density, diffusive coalescence mechanism, due to Binder and Stuffer, is commonly used \cite{binder_stauffer}. However, in the considered situations, due to inviscid vapor background the clusters of varying fractal dimensions move ballistically. A ballistic aggregation theory \cite{carnevale,trizac2003correlations,midya2017prl,ulrich2009dilute,pathak2014inhomogeneous,trizac1995dynamic,paul2017ballistic} nicely describes the simulation data. It appears that the theory works for a wide range of parameter values. Furthermore, the theory also suggests exponential growth. While this is in deviation from the standard expectation of phase ordering dynamics, such situation should be achievable in real experiments. 

Both simulation and theoretical outcomes of this study have important implications in a wide variety of physical situations including structure formation in cosmic dust and aerosols. Ballistic aggregation has been a topic of much theoretical interest in studies of granular matter and elsewhere \cite{carnevale,trizac1995dynamic,paul2017ballistic,cremer2014scaling,puthalath2023lattice,venkatareddy2025phase,bsgupta2026active,paul2021clusters,midya2017prl}. In some of these studies models were constructed to directly produce the ballistic motion and fractality. In our study, with a realistic model, these features appear naturally, mimicking direct experimental conditions of phase transitions.

%
\end{document}